\documentclass[journal]{IEEEtran}
\usepackage{tikz}
\usetikzlibrary{external}
\usepackage[caption=false,font=footnotesize]{subfig}
\usepackage{blindtext}
\usepackage{graphicx}
\usepackage{float}
\usepackage{cite}
\usepackage{amsmath}
\usepackage{amssymb}
\usepackage{algorithmic}
\usepackage{stfloats}
\usepackage{makecell}
\usepackage{mathtools}
\usepackage[hidelinks]{hyperref}
\usepackage{threeparttable}
\usepackage[bottom]{footmisc}
\usepackage[T1]{fontenc}
\usepackage{lscape}
\usepackage{wrapfig}
\usepackage{capt-of}
\usepackage{booktabs}
\usepackage{varwidth}
\usepackage{array}
\usepackage{ragged2e}
\usepackage{xcolor} 

\definecolor{color0}{rgb}{0.12156862745098,0.466666666666667,0.705882352941177}
\definecolor{color1}{rgb}{1,0.498039215686275,0.0549019607843137}
\definecolor{color2}{rgb}{0.172549019607843,0.627450980392157,0.172549019607843}
\definecolor{color3}{rgb}{0.83921568627451,0.152941176470588,0.156862745098039}
\definecolor{color4}{rgb}{0.580392156862745,0.403921568627451,0.741176470588235}
\definecolor{color5}{rgb}{0.549019607843137,0.337254901960784,0.294117647058824}
\definecolor{color6}{rgb}{0.890196078431372,0.466666666666667,0.76078431372549}

\usepackage{booktabs}
\usepackage{colortbl}
\usepackage{tabularx}
\newcolumntype{o}{>{\hsize=4.2\hsize}X}
\newcolumntype{t}{>{\hsize=5.0\hsize}X}
\newcolumntype{F}{>{\hsize=4.2\hsize\raggedleft\arraybackslash}X}
\usepackage{tabulary}
\newcolumntype{I}{>{\hsize=1.5\hsize}J}
\usepackage{multicol} 
\usepackage{enumitem}
\setlist[itemize]{nosep}
\setlist[enumerate]{nosep}
\definecolor{orange}{rgb}{1,0.5,0}
\definecolor{light-gray}{gray}{0.95}

\usepackage{pgfplots}
\pgfplotsset{compat=1.16}
\usepgfplotslibrary{groupplots,dateplot}
\usepackage[american,cuteinductors,smartlabels, RPvoltages]{circuitikz}
\usetikzlibrary{babel}
\usetikzlibrary{calc, arrows, arrows.meta, patterns,shapes.arrows, positioning, plotmarks, matrix}
\ctikzset{bipoles/thickness=1}
\ctikzset{bipoles/length=0.8cm}
\ctikzset{bipoles/diode/height=.375}
\ctikzset{bipoles/diode/width=.3}
\ctikzset{tripoles/thyristor/height=.8}
\ctikzset{tripoles/thyristor/width=1}
\ctikzset{bipoles/vsourceam/height/.initial=.7}
\ctikzset{bipoles/vsourceam/width/.initial=.7}
\tikzstyle{every node}=[font=\small]
\tikzstyle{every path}=[line width=0.8pt,line cap=round,line join=round]
\definecolor{AleeRed}{rgb}{0.5,0,0}

\tikzset
{A/.style={color=blue, 
line width = 3pt,
},
B/.style={color=red,
},
C/.style={color=green!60!black,
}
}
\pgfplotsset{mystyle/.style={%
    width=8.25cm,
    height=6.0cm,
    xmin=-2,xmax=3,
    xtick={-2,-1,...,3},
    minor x tick num=1,
    minor y tick num=1,
    major tick length=0.15cm,
    minor tick length=0.075cm,
    color=black, 
    line width = 5pt,
    axis line style = very thick,
    xmajorgrids,
    ymajorgrids,
    legend style={
        fill opacity=0.8,
        draw opacity=1,
        text opacity=1,
        line width = 1 pt,
        at={(0.03,0.97)},
        anchor=north west,
        draw=white!80!black
    },
    grid style = {
        dash pattern = on 0.05mm off 1mm,
        line cap = round,
        black,
        line width = 0.65pt
        },
    }
}
\pgfmathsetlengthmacro\MajorTickLength{
      \pgfkeysvalueof{/pgfplots/major tick length} * 0.6
    }
\pgfplotsset{basic/.style={%
    width=8.5cm,
    height=6.0cm,
    xlabel={Eggs (no.)},
    color=black, 
    line width = 5pt,
    axis line style = very thick,
    legend style={
        fill opacity=0.8,
        at={(0.03,0.97)},
        anchor=north west,
    },
    major tick length=\MajorTickLength,
      every tick/.style={
        black,
        thick,
      },
    xmajorgrids,
    ymajorgrids,
    grid style = {
        dotted,
        line cap = round,
        black,
        line width = 0.40pt,
        opacity=0.3
        },
    }
}
\pgfmathdeclarefunction{minEntropyRight}{1}{%
  \pgfmathparse{-(log2(x)}%
}
\pgfmathdeclarefunction{minEntropyLeft}{1}{%
  \pgfmathparse{-(log2(1-x)}%
}
\pgfmathdeclarefunction{shannonEntropy}{1}{%
  \pgfmathparse{-(x*log2(x)+(1-x)*log2(1-x))}%
}
\usepackage{listings} 
\lstdefinelanguage{vim}
{
  morekeywords={
  set, let
  map, nmap,
  filetype,
  on, off,
  autocmd,
  Plugin,it
  call,
   },
morecomment=[l]{"}, 
morestring=[b]' 
}

\newcommand\unaryminus{\smash{\scalebox{0.5}[1.0]{\( - \)}}}

\tikzset{
    plot box/.style={
        red, dashed,
        thick
    },
    plot box padding/.initial=5pt
}

\def\normalcoord(#1){coordinate(#1)}
\def\showcoord(#1){coordinate(#1) node[circle, red, draw, inner sep=1pt,
pin={[red, overlay, inner sep=0.5pt, font=\tiny, pin distance=0.1cm,
pin edge={red, overlay}]45:#1}](){}}

\makeatletter
\def\grd@save@target#1{%
  \def\grd@target{#1}}
\def\grd@save@start#1{%
  \def\grd@start{#1}}
\tikzset{
  grid with coordinates/.style={
    to path={%
      \pgfextra{%
        \edef\grd@@target{(\tikztotarget)}%
        \tikz@scan@one@point\grd@save@target\grd@@target\relax
        \edef\grd@@start{(\tikztostart)}%
        \tikz@scan@one@point\grd@save@start\grd@@start\relax
        \draw[minor help lines] (\tikztostart) grid (\tikztotarget);
        \draw[major help lines] (\tikztostart) grid (\tikztotarget);
        \grd@start
        \pgfmathsetmacro{\grd@xa}{\the\pgf@x/1cm}
        \pgfmathsetmacro{\grd@ya}{\the\pgf@y/1cm}
        \grd@target
        \pgfmathsetmacro{\grd@xb}{\the\pgf@x/1cm}
        \pgfmathsetmacro{\grd@yb}{\the\pgf@y/1cm}
        \pgfmathsetmacro{\grd@xc}{\grd@xa + \pgfkeysvalueof{/tikz/grid with coordinates/major step}}
        \pgfmathsetmacro{\grd@yc}{\grd@ya + \pgfkeysvalueof{/tikz/grid with coordinates/major step}}
        \foreach \x in {\grd@xa,\grd@xc,...,\grd@xb}
        \node[anchor=north] at (\x,\grd@ya) {\pgfmathprintnumber{\x}};
        \foreach \y in {\grd@ya,\grd@yc,...,\grd@yb}
        \node[anchor=east] at (\grd@xa,\y) {\pgfmathprintnumber{\y}};
      }
    }
  },
  minor help lines/.style={
    help lines,
    opacity=.25,
    step=\pgfkeysvalueof{/tikz/grid with coordinates/minor step}
  },
  major help lines/.style={
    help lines,
    opacity=.35,
    line width=\pgfkeysvalueof{/tikz/grid with coordinates/major line width},
    step=\pgfkeysvalueof{/tikz/grid with coordinates/major step}
  },
  grid with coordinates/.cd,
  minor step/.initial=.2,
  major step/.initial=1,
  major line width/.initial=1pt,
}
\makeatother
\usetikzlibrary{shadows.blur}
\usetikzlibrary{shapes.symbols}
\pgfdeclareverticalshading{pgf@lib@fade@north}{100bp}
{color(0bp)=(pgftransparent!0);
color(25bp)=(pgftransparent!0);
color(75bp)=(pgftransparent!100);
color(100bp)=(pgftransparent!100)}%
\usetikzlibrary{fadings}
\makeatletter
\pgfdeclareverticalshading{pgf@lib@fade@north}{100bp}
{color(0bp)=(pgftransparent!0);
 color(5bp)=(pgftransparent!10);
 color(60bp)=(pgftransparent!100); color(70bp)=(pgftransparent!100)}%
\pgfdeclarefading{myfade}{%
  \pgfuseshading{pgf@lib@fade@north}%
}
\makeatother

\usetikzlibrary{fit}

\begin{document}
\title{A 3.3 Gbps SPAD-Based Quantum Random Number Generator}
\author{Pouyan~Keshavarzian,~\IEEEmembership{Student~Member,~IEEE,}
        Karthick Ramu,
        Duy Tang,
        Carlos Weill,
        Francesco~Gramuglia,~\IEEEmembership{Student~Member,~IEEE,}
        Shyue~Seng~Tan,
        Michelle~Tng,
        Louis Lim,
        Elgin~Quek,~\IEEEmembership{Member,~IEEE,},
        Denis Mandich,
        Mario Stipčević,
        and~Edoardo~Charbon,~\IEEEmembership{Fellow,~IEEE}
\thanks{}
\thanks{This work was supported by the Swiss National Science Foundation under Grant  200021-169465. The work of Pouyan Keshavarzian was supported by Qrypt Inc.}
\thanks{This work has been submitted to the IEEE for possible publication. Copyright may be transferred without notice, after which this version may no longer be accessible.}
\thanks{}
\thanks{}
\thanks{}
\thanks{}
}

\markboth{}%
{}

\maketitle
\begin{abstract}
Quantum random number generators are a burgeoning technology used for a variety of applications, including modern security and encryption systems. Typical methods exploit an entropy source combined with an extraction or bit generation circuit in order to produce a random string. In integrated designs there is often little modelling or analytical description of the entropy source, circuit extraction and post-processing provided. In this work, we first discuss theory on the quantum random flip-flop (QRFF), which elucidates the role of circuit imperfections that manifest themselves in bias and correlation. Then, a Verilog-AMS model is developed in order to validate the analytical model in simulation. A novel transistor implementation of the QRFF circuit is presented, which enables compensation of the degradation in entropy inherent to the finite non-symmetric transitions of the random flip-flop. Finally, a full system containing two independent arrays of the QRFF circuit is manufactured and tested in a 55 nm Bipolar-CMOS-DMOS (BCD) technology node, demonstrating bit generation statistics that are commensurate to the developed model. The full chip is able to generate 3.3 Gbps of data when operated with an external LED, whereas an individual QRFF can generate 25 Mbps each of random data while maintaining a Shannon entropy bound > 0.997, which is one of the highest per pixel bit generation rates to date. NIST STS is used to benchmark the generated bit strings, thereby validating the QRFF circuit as an excellent candidate for fully-integrated QRNGs. 
\end{abstract}

\section{Introduction}
\IEEEPARstart{R}{andom} number generators (RNGs) are well-established security primitives used in a variety of schemes ranging from key generation/distribution to, encryption, and privacy amplification \cite{Alioto_basicsAsics_SSC2019}. With the proliferation of the Internet of Things (IoT) and connected devices, security has become a critical aspect of all system-level design. Consequently, true random number generators (TRNGs) \cite{Satpathy_SOA_TRNG_2019,Bae_SOA2_TRNG_2019}, which exploit some classical physical entropy source, are a mature technology available commercially as both discrete silicon devices and IP blocks inside more complex computing circuitry \cite{VonKaeknel_TRNGCPU_2007} and are able to achieve energy per bit ratios lower than pJ/bit \cite{Larimian_lightweightTRNG_2020}. However, due to the inherent limitations of classical entropy sources in providing sufficient randomness, i.e. limitation of bit bias and correlation, these TRNG ASICs often require complex post-processing in order to establish an acceptable output entropy in the generated bit stream, which in turn significantly reduces the bit rate output \cite{Alioto_basicsAsics_SSC2019}. Finally with the emergence of quantum computing, the required security parameter for a generated key increases, doubling the required key length for symmetric encryption algorithms \cite{Sabah_RoleOfHashBasedSignatures_2021,SecureIoTQuantum_Cheng_2017}. 

Quantum random number generators, which exploit inherently random phenomena in nature, are promising technologies which aim to address the challenge/tradeoff between system complexity and randomness performance. QRNG standardisation is underway while debate remains regarding requirements for and specifics of post-processing methods \cite{Ma_QRNGPostProc_2013,Rozic_TRNGPostP_2019}, along with the validity of randomness testing \cite{HurleySmith_QLeapCreash_ACM2020,Kinga_NISTinterp_SciAndTech2015}. Nevertheless, the exploitation of quantum phenomena provides advantages for the development of future random number generators, particularly for Entropy-as-a-Service (EaaS) and quantum key distribution (QKD) applications, which necessitate very high bit generation rates. 

Systems and methods for QRNG designs come in many flavors, including those which exploit photon timing statistics, polarization, quantum tunneling, laser phase noise, to name a few. Furthermore, as an additional measure for combating environmental changes or attacks on the device itself, complex generators that are proven to be device \cite{Cao_SourceIndQRNG_2016} and measurement independent\cite{Cao_MeasuIndQRNG_2015} have been demonstrated in literature, although they remain very impractical owing to the very low bit rate (bits-kbps) and bulky setups. A compromise between so-called trusted systems that suggest the quantum nature of the entropy source can create a sufficient generator, and those that contrive more secure bounds, using post-processing or source/device independence, are so-called self-testing quantum random number generators that test for generator defectiveness \cite{Rusca_selfTestQRNG_PhysRevA2019,Lunghi_selfTestQRNG_PhysRevA2015}. This is performed by creating tests tailored specifically to verify the generator output string against its randomness model. Regardless of the generator design, those which provide the most pragmatic solution can be readily modelled, integrated in silicon, and scalable in order to produce designs with high data throughput. For these reasons, single-photon avalanche diode (SPAD) based systems are attractive for QRNG technology development as they are highly scalable ($>$ 1 Mpixel \cite{morimoto2020megapixel,morimorotCanan3p2Meg_IEDM2021}) and reproducible in silicon anianufacturing. 

The composition of this paper is as follows. In Section II, we review some previously developed theory on the quantum random flip-flop (QRFF) circuit \cite{Stipcevic_QRFF_2016,stipcevic2021scalable} and thereby introduce the considerations for an integrated circuit that employs this method. From there, a Verilog-AMS model (Section III) is developed in order to thoroughly investigate, in simulation, how circuit imperfections manifest themselves in bias and correlations, thereby validating the analytical model of the bit generation method. In Section VI, a novel full-custom implementation of the QRFF flip-flop is proposed, which uses dynamic logic to overcome effects of finite and non-symmetric transitions present in logic circuits, on the quality of generated bit strings. This QRFF is then implemented in a 55 nm BCD process with measurements comparing the results to the analytical and simulated predictions provided. Finally, we scale the QRFF circuit to a full Gbps QRNG design on chip. Two independent arrays, capable of running concurrently, are implemented with separate readout schemes and achieve a combined 3.3 Gbps output data-rate showing the suitability of the approach in practice. A discussion is provided in Section VII, followed by a conclusion in Section VIII 
   
\section{The Quantum Random Flip-Flop}
\begin{figure*}[!t]
\centering
\subfloat[Circuit symbol.]{
\includegraphics[width=0.2\textwidth]{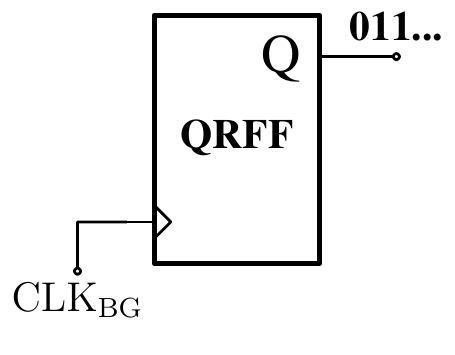}
\label{fig:qrffsymbol}}\hspace{-8mm}
\subfloat[qrff circuit implmentation]{
\includegraphics[width=0.33\textwidth]{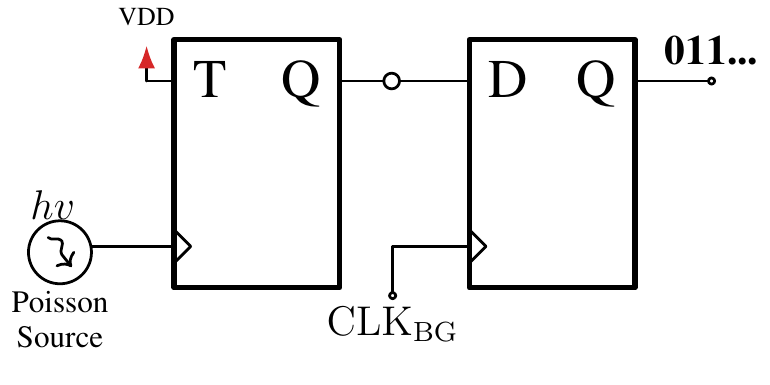}
\label{fig:qrff_implmentation}}\hspace{-2mm}
\subfloat[Realistic waveform.]{
\includegraphics[width=0.35\textwidth]{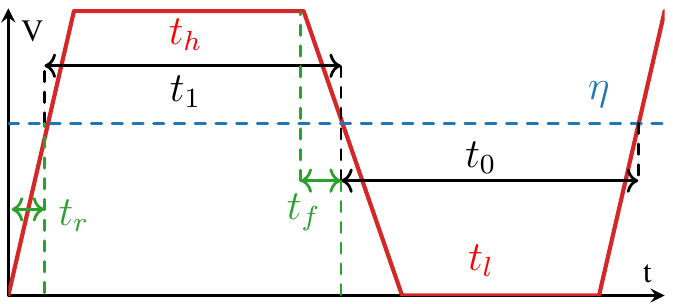}
\label{fig:qrrwaveform}}
\caption{An example circuit implementation of the QRFF concept presented in this work with a TFF and DFF combined with an exponentially distributed counting source. The waveform of the TFF output is shown to illustrate pertinent characteristics such as the normalized sampling threshold ($\eta$), rise and fall times ($t_r$, $t_f$) that contribute to bias.}
\label{fig:qrff}
\end{figure*}
\subsection{Fundamental Operation}
The QRFF describes a simple circuit concept that, upon the arrival of a clock strobe, generates a random bit. A symbol representation is shown in \figurename~\ref{fig:qrffsymbol}. A specific circuit realization of the QRFF concept is shown in \figurename~\ref{fig:qrff_implmentation}. Here, an exponentially distributed (in time) discrete event clocks a toggle flip-flop that has its toggle input continuously held to logic 1, thereby realizing the well-known random telegraph signal/process (RTS) with random transitions in time. In principle, as the arrival events occur randomly, the TFF output, over a sufficient integration, is uniformly distributed $X\sim U\{0,1\}$. Therefore, once the sampling DFF is clocked by the strobe signal $\mathrm{CLK_{BG}}$, a random bit is generated. The architectural simplicity allows for accurate modeling of the bias and autocorrelation of generated bit strings, while the ability to vary internal parameters such as the arrival rate of Poisson events, and external parameters, such as the generation rate, enables flexibility from a system point-of-view, which we will demonstrate in further sections. 

\subsection{Model for bias and correlation}
Evidently, no imperfectionless source or circuit can exist which then perfectly matches the theory of the \figurename~\ref{fig:qrffsymbol} concept. The output of the TFF indeed has finite, and non-symmetric rise/fall times. Furthermore, the sampling threshold of the signal, which distinguishes between low and high has some deviation from center, resulting in a RTS that resembles the waveform depicted in \figurename~\ref{fig:qrrwaveform}. The time between transition edges, $\tau_{\mathrm{D}}$, is determined by the detection rate $\lambda_{\mathrm{D}}=1\mathbin{/}\tau_{\mathrm{D}}$ and splits the signal into two equal half-periods. The rise and fall times are denoted by $t_r$ and $t_f$, respectively, and represent the transition time between the `1' and `0' states until the level of the normalized sampling threshold, $\eta$. 

It can be shown that the statistical bias i.e. deviation from $\mathrm{P}(X=1)=0.5$ for the high state is described by (\ref{eqn:qrffbiaseqn}) \cite{stipcevic2021scalable}. \vspace{1.5mm} 
\begin{equation}
b = \dfrac{t_f-\eta(t_r+t_f)}{2}\cdot\lambda{_\mathrm{D}}. \label{eqn:qrffbiaseqn}
\vspace{1.5mm}
\end{equation}
Some key guidelines for circuit design can be extracted from this model. First, it is clear that the bias should scale linearly in magnitude with increasing detections and that it is desirable to have a fast TFF. Furthermore, it should be possible to essentially eliminate bias resulting from any non-symmetry of the rise/fall times by dynamically adjusting the sampling threshold. 

Sources of correlation in any RNG must also be modelled and understood. The autocorrelation function for a binary RTS with normalized amplitudes is defined by (\ref{eqn:qrrcorreqn}). \vspace{1.5mm} 
\begin{equation}
\mathrm{R}_{XX} (\tau) = e^{\left(-2\cdot\lambda\cdot|\tau|\right)}. \label{eqn:qrrcorreqn}
\vspace{1.5mm}
\end{equation}  
The time lag interval, $\tau$, for calculation of the autocorrelation coefficient, is controlled by the clock frequency of the sampling DFF in \figurename~\ref{fig:qrff_implmentation}. Therefore, correlation coefficients, $a_i$, corresponding to specific bit lags, $i$, can be calculated with (\ref{eqn:qrrcorrcoeffeqn}). \vspace{1.5mm}
\begin{equation}
a_{i} = e^{\left(-2\cdot\lambda_{\mathrm{D}}/\left(i\cdot f_{\mathrm{BG}}\right)\right)}. \label{eqn:qrrcorrcoeffeqn}
\vspace{1.5mm}
\end{equation}

The 1-bit lag correlation coefficient, $a_1$, therefore has the highest magnitude, and can be minimized by increasing the ratio ${\lambda_{\mathrm{D}}}/{f_{\mathrm{BG}}}$. Consequently, there exists an inherent tradeoff between designing for acceptable bias, which increases linearly, and for correlation, which decreases exponentially, with increasing detection rate. 

This model is limited by the assumptions that a true Poisson counting process is used as the entropy source. Therefore, care must be taken to ensure that detector imperfections, such as afterpulsing, are negligible or reduced to a minimum, and that the circuit/illumination conditions allow for consistent detection dead times.

\subsection{Benchmarks for Performance}
As noted earlier, while the security requirements of any given system and cryptographic scheme can vary, we aim to design a generator which is capable of complying with entropy requirements of the AIS-31 standard,therefore, the Shannon entropy, $\mathrm{H}_1(X)= \unaryminus\sum_{i=1}^{n}p_i\log p_i$, must remain $\geq$ 0.997 for a sufficiently long bit string \cite{ais31}. The corresponding acceptable bias and correlation values must therefore remain below a level of $10^{\unaryminus 3}$. The NIST Statistical Test Suite is used to validate the performance of overall bit strings generated by the final array.  
\section{Verilog-AMS Simulation of QRFF Analytical Model}
\subsection{Model details}
In order to validate these analytical equations, a simple SPICE-compatible Verilog-AMS model of the QRFF circuit was developed. An exponential source was used by taking advantage of the \textit{\$rdist\_exponential} function provided by the Verilog-AMS language standard. The parameters in Table~\ref{tab:qrffsimparams} were investigated as variables in simulation of bias, $b$, and correlation coefficients, $a_i$.
\begin{figure}[!t]
\centering
\subfloat[Rise/fall time ($t_r,t_f$) mismatch analysis: with fixed detection rate  ($\lambda{_\mathrm{D}}=$ 80 Mcps), bit generation rate ($f_{\mathrm{BG}} = 25$ MHz and normalized sampling threshold ($\eta=0.499$).]{
\includegraphics[]{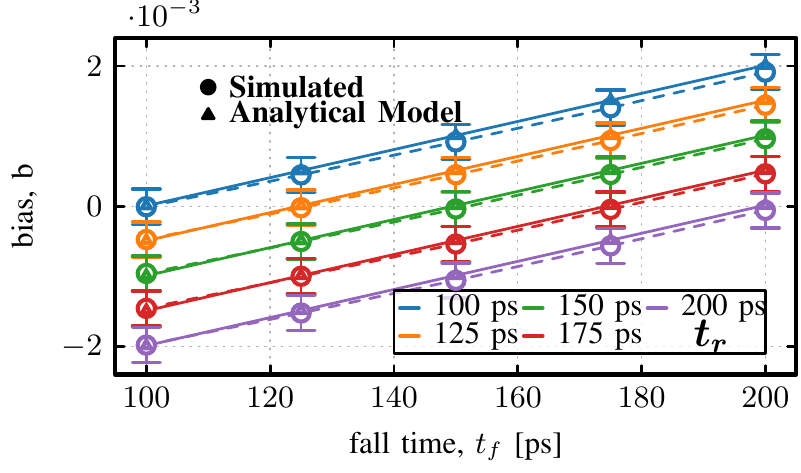}
\label{fig:biassimtrtfsweep}}\\
\subfloat[Sampling threshold analysis ($\eta$): fixed detection rate ($\lambda{_\mathrm{D}}=$ 80 Mcps), and bit generation rate ($f_{\mathrm{BG}} = 25$ MHz) performed at various TFF rise/fall.]{
\includegraphics[]{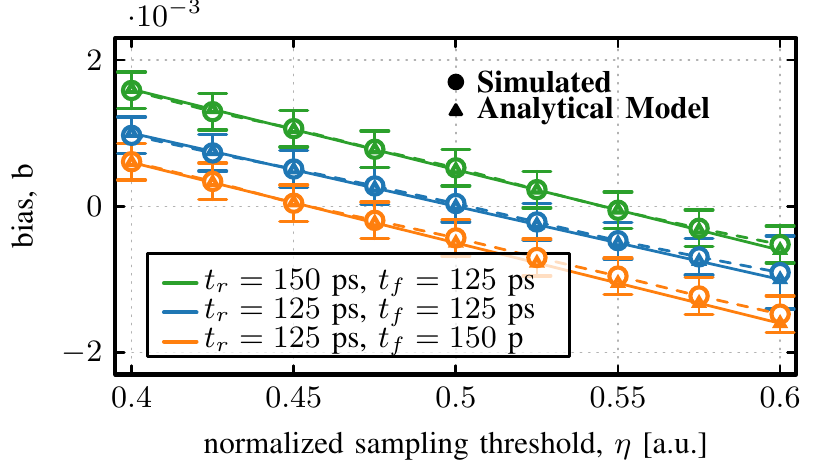}
\label{fig:biassimvtsweep}}\\
\subfloat[Detection rate ($\lambda{_\mathrm{D}}$) analysis with: fixed normalized sampling flip-flop threshold ($\eta=0.475$), fixed bit generation rate ($f_{\mathrm{BG}} = 25$ MHz). Rise/fall time discrepancy is deliberately exaggerated in order to increase bias so the number of samples simulated can be reduced and still be statistically relevant.]{
\includegraphics[]{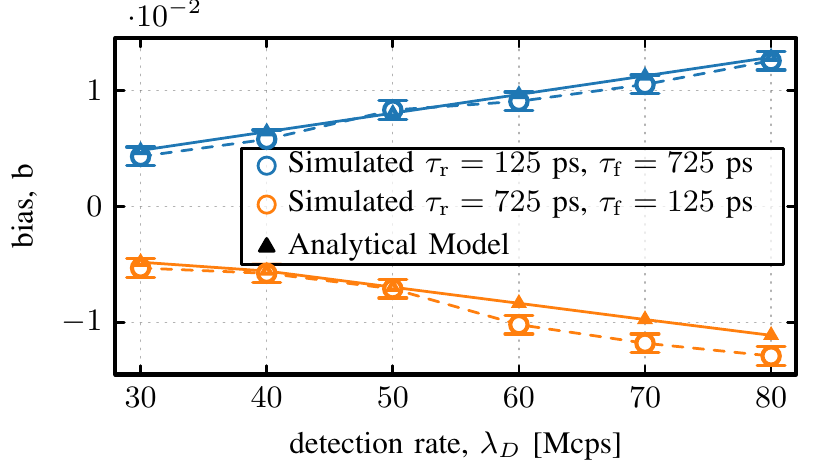}
\label{fig:biassimcountsweep}}
\vfill
\caption{$P=1$ bias analysis simulations with comparison to the analytical model (\ref{eqn:qrffbiaseqn}). $\sigma$ for each simulation is plotted as error bars.}
\label{fig:simbiasanal}
\end{figure}
\begin{table}[!ht]
\caption{Verilog-A Model Parameters of QRFF}
\centering
\begin{tabular}{|c||l|}
\hline
\textbf{Param.} &  \textbf{Description} \\ \hline\hline
 $\eta$ & DFF sampling threshold normalized to 1 V \\ \hline 
 $t_r$ & TFF rise time \\ \hline
 $t_f$ & TFF fall time \\ \hline
 $\lambda_{\mathrm{D}}$ &  Detection rate from exponential source\\ \hline
 $f_{\mathrm{BG}}$ & Sampling frequency of bit gen. clock ($\mathrm{CLK_{BG}}$)\\\hline
\end{tabular}\label{tab:qrffsimparams}
\end{table}
\subsection{Simulation results}
Simulation results of bias are displayed in \figurename~\ref{fig:simbiasanal}. The generation of bits is a Binomial process with $N$ trials, therefore the variance of bias from simulation can be calculated with $\sigma^2=1/(4N)$. In our results, we plot $\pm\sigma$ for reference. \figurename~\ref{fig:biassimtrtfsweep} displays the simulated bias compared to the analytical calculation for varying $t_r,t_f$, given a fixed detection rate $\lambda_\mathrm{D}=80$ Mcps and and a sampling threshold, $\eta= 0.499$, placed close to the center of the waveform. As the discrepancy between the rise and fall time increases, so does bias, matching very closely to the analytical calculation. In \figurename~\ref{fig:biassimvtsweep}, a similar analysis was performed but with a varying $\eta$. Here, we can see that a mismatch between $t_r$ and $t_f$ can be compensated for by adjusting the threshold, thereby allowing for the minimization of bias. This is a critical finding from the perspective of integrated circuit implementation, as the foundry process will always create some small, albeit present, variation, across an array regardless of how carefully the circuit is designed. In order to confirm that the bias magnitude scales linearly with increased count rate, at a fixed sampling rate and threshold, a final simulation is performed with the results displayed in \figurename~\ref{fig:biassimcountsweep}.

\begin{figure}[!t]
\centering
\subfloat[1-bit lag autocorrelation analysis with fixed normalized sampling flip-flop threshold ($\eta=0.475$) at $f_{\mathrm{BG}} = 25$ MHz and swept across detection rates ($\lambda{_\mathrm{D}}$).]{
\includegraphics[]{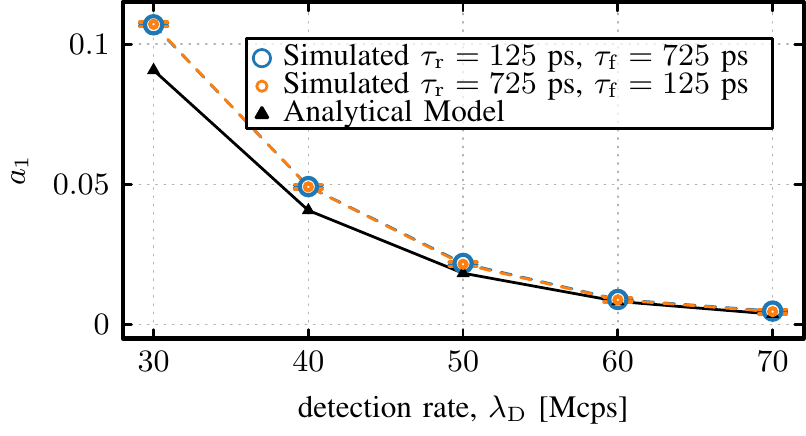}
\label{fig:a1simcountsweep}}\\
\subfloat[Autocorrelation analyses with lags of 1-3 bits, fixed normalized sampling flip-flop threshold: $\eta=0.475$, fixed TFF rise/fall times: $\tau_r = 725$ ps and $\tau_f = 125$ ps, and fixed detection rate: $\lambda{_\mathrm{D}}=40$ Mcps.]{
\includegraphics[]{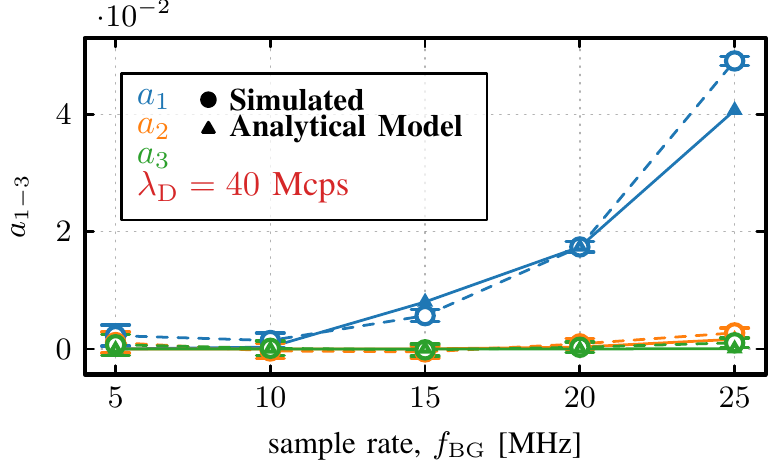}
\label{fig:a1-3_sim_fbgsweep}}\\
\subfloat[$P=1$ bias and 1-bit lag autocorrelation analyses with fixed normalized sampling flip-flop threshold: $\eta=0.475$, fixed TFF rise/fall times: $\tau_r = 725$ ps and $\tau_f = 125$ ps, and fixed detection rate: $\lambda{_\mathrm{D}}=40$ Mcps.]{
\includegraphics[]{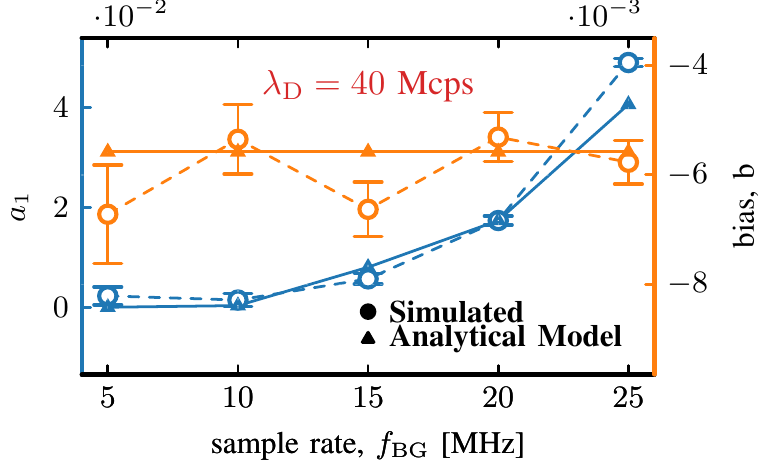}
\label{fig:biasanda1_sweptfbg}}
\caption{Spatial bias map across excess bias.}
\label{fig:autocorrsims}
\end{figure}

Autocorrelation simulations with comparison to the calculation of an RTS, eqaution (\ref{eqn:qrrcorrcoeffeqn}), are presented in \figurename~\ref{fig:autocorrsims}. At a fixed sampling rate, the 1-bit lag correlation coefficient should decrease exponentially, which is indeed observed in the results of \figurename~\ref{fig:a1simcountsweep}. Conversely, at a fixed detection rate, the correlation should increase exponentially for increased sample rates, as shown in \figurename~\ref{fig:a1-3_sim_fbgsweep}. Although, as predicted, higher order coefficients remain very low. The modelling suggests that, given a constant detection rate and circuit speed parameters, the bias should remain unchanged with varied sampling rates. To demonstrate this, the simulated data for $a_1$ in \figurename~\ref{fig:a1simcountsweep} is plotted once more, along with the bias, in \figurename~\ref{fig:biasanda1_sweptfbg}. The length of the simulation for each data point was kept constant, therefore the total number of generated bits vary. For this reason, the $\sigma$ increases with decreased sample rate. 

Some relevant system considerations can be derived from the above analysis. Given a detection rate of 80 Mcps, which is readily achievable in an integrated SPAD circuit, and a DFF which contains a tunable sampling threshold, a generator which is capable of producing 25 Mbps per pixel is achievable while maintaining a Shannon entropy bound of 0.997. This per pixel generation rate is considerably higher than those demonstrated by other SPAD-based QRNG techniques \cite{Massari_arbiterQRNG_ISSCC2016, Tisa_MilanQRNG_JSTQE2015,Massari_arbiterQRNG_ISSCC2016}. Finally, the model could be further improved by formulating the effects of detector imperfections, in particular those containing correlated effects, such as afterpulsing and crosstalk. Clearly, this analysis is only effective for a single QRFF, therefore exploration of system consideration such as PVT of the TFF, count-rate/breakdown non-uniformity, and others, must be performed in order to have a clear view of the scalability of this circuit concept. 

\section{Design of a full-custom CMOS QRFF}
\begin{figure*}[!t]
    \centering
    \includegraphics[]{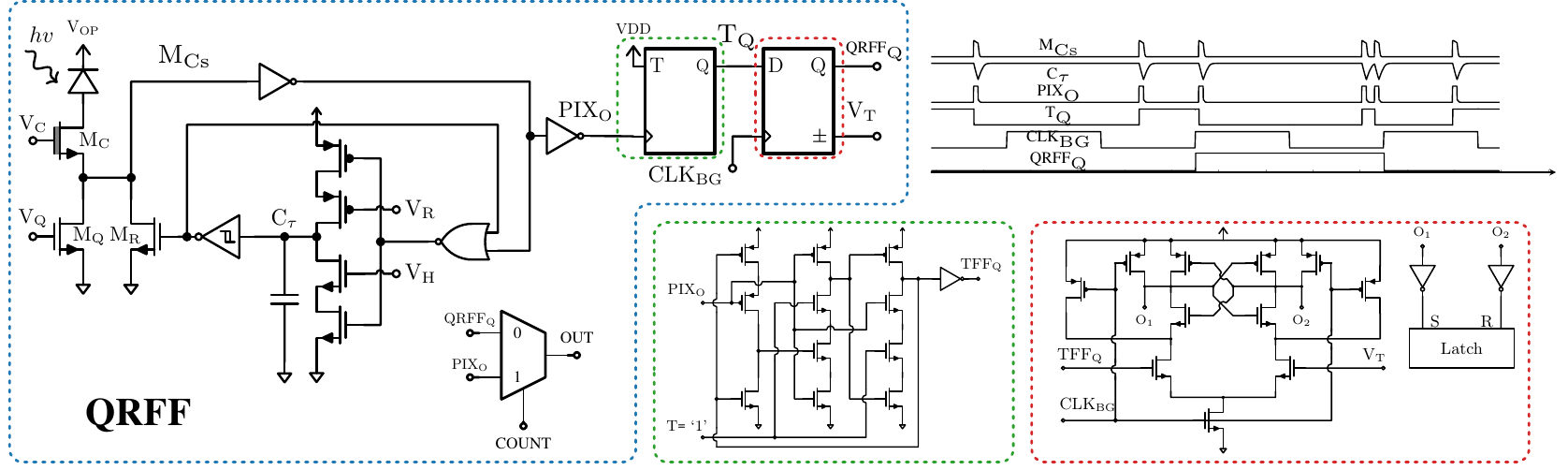}
    \caption{Complete custom QRFF design including PQAR circuit and full custom flip-flop design for improved performance.}
    \label{fig:pix}
\end{figure*} 
\subsection{Technology Consideration}
Recently, it has been shown that low afterpulsing detectors (< 1\%) are achievable in multiple deep sub-micron processes \cite{PellST40_IndustrailSPADs_IEDM2017,GramugliaKesharzianKizilkan_Eng55nm_JSTQE2022}. Moreover, high-brightness MicroLEDs \cite{XuMIT_LED_2021} and SPADs have both been demonstrated in the 55 nm BCD process. These recent advances bring new promise for research into the commercially-viable SPAD-based QRNG sensors, especially for architectures which employ the QRFF method. In this work, we take a step towards that vision by integrating all the detection, bit generation and readout circuitry while leaving the illumination external.     
\subsection{Pixel Design}
In order to test the model presented, and take advantage of the findings from the simulation analysis, which demonstrates the ability to overcome circuit imperfections, a pixel design containing a full-custom version of the QRFF is proposed and shown in \figurename~\ref{fig:pix}. Although very-high performing SPADs were recently demonstrated in the GF 55 nm BCD process \cite{GramugliaKesharzianKizilkan_Eng55nm_JSTQE2022}, it is not considered a mature CMOS image sensing process, as a standard flow was used for the fabrication of this chip. Therefore, several tunable pixel functions were implemented in order to limit detector variability. 

For the TFF, a true-single-phase clock (TSPC) logic-based circuit was implemented for enabling fast transitions, with the output buffer sized appropriately for symmetric rise/fall times. However, as previously stated, process variation will always result in some mismatch across the array. For this reason, a comparator based sampling flip-flop is an evident choice in order to achieve a mean bias centered at zero, overcoming any inevitable non-symmetry. A strongARM comparator-based DFF was designed for fast latching, further enabling high-speed solution which require serialization of many QRFF's onto a readout bus. The sampling threshold of the DFF is controlled by a global signal $\mathrm{V_T}$.

The pixel employs a passive-quench active-recharge (PQAR) circuit in order to limit afterpulsing based off of the design from \cite{keshavarzian_APCXVI_2022}. The passive-quench transistor, $\mathrm{M_Q}$, is designed for a high-impedance, limiting charge flow, which reduces the population of trapped carriers upon an avalanche \cite{BronziFastQQuenchPixe_2013}, and quickly quenches the SPAD. Reduction of the SPAD bias, $\mathrm{V_{OP}}$, also reduces afterpulsing. However, since the variability of breakdown voltages in this process, until this point in time, remained unexplored, it was critical to allow for large range of excess bias values so that all pixels in the array can be utilized. For this reason, a thick-oxide cascode transistor, $\mathrm{M_C}$, was chosen so that higher excess bias values can be used without damaging the electronics. A voltage-controlled tunable delay element in the monostable feedback loop was implemented to further investigate the optimal dead-time i.e. a high count rate/afterpulsing tradeoff. The hold and recharge times of the SPAD pulse are determined by the discharging and recharging time of the feedback capacitor, $\mathrm{C_\tau}$, which can be adjusted using the global control pins, $\mathrm{V_H}$ and $\mathrm{V_R}$. As $\mathrm{V_H}$ is increased, the discharging time of $\mathrm{C_\tau}$ decreases, thereby decreasing the length of time until $\mathrm{M_R}$ is turned on following an avalanche, consequently decreasing the hold time. Conversely, increasing of $\mathrm{V_R}$ adjusts the length of time for which $\mathrm{M_R}$ is on, allowing for a controllable recharge time.    

This complete pixel, represents a realization of a QRFF, and its general functionality is described by the timing diagram in \figurename~\ref{fig:pix}. Upon an avalanche detection, the SPAD becomes inactive until recharged, which is determined by the external voltage control, and the TFF is consequently toggled. With the arrival of the global bit generation clock signal, $\mathrm{CLK_{BG}}$, a random bit is generated at the output, $\mathrm{QRFF_Q}$. 

\section{QRNG Architecture and Characterization Setup}
\begin{figure}[!ht]
    \centering
    \includegraphics[]{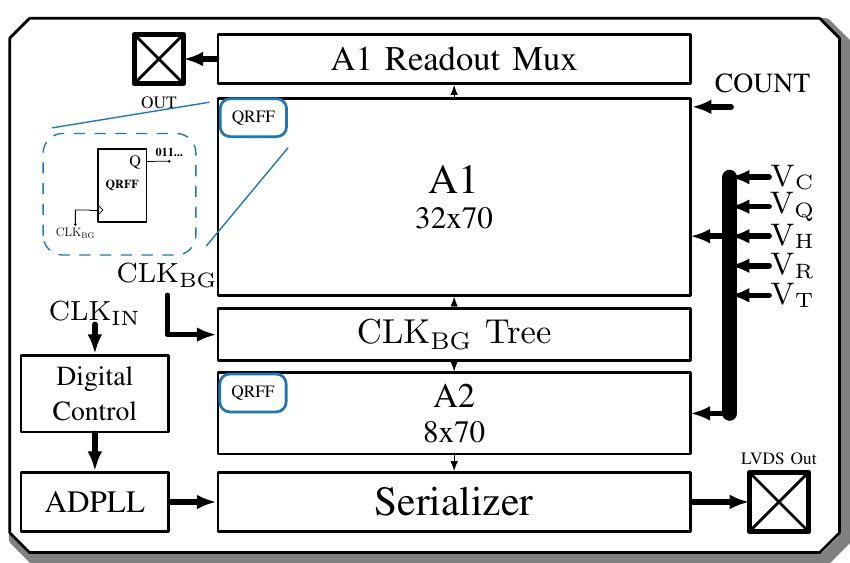}
    \caption{FortunaSPAD full block diagram.}
    \label{fig:fortunablock}
\end{figure}
\begin{figure}[!ht]
    \centering
    \includegraphics[]{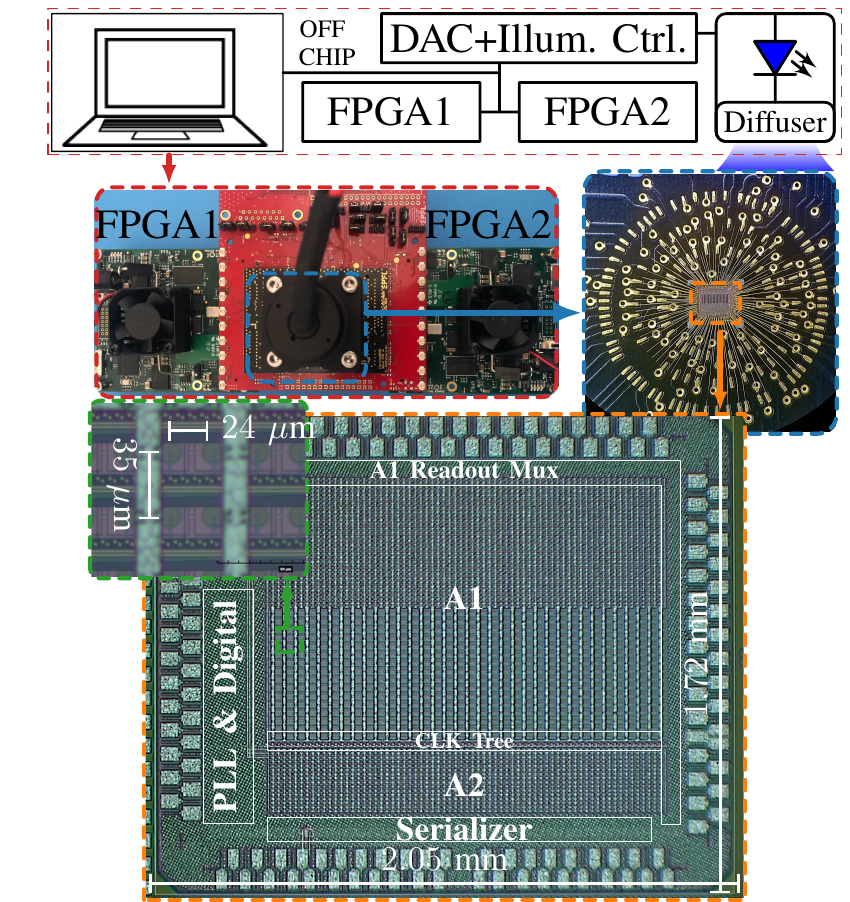}
    \caption{FortunaSPAD micrograph with characterization setup including readout/control FPGAs. The die total die are is 2.05mm x 1.72mm.}
    \label{fig:chiparchandtest}
\end{figure} 
A sensor with 2800 total QRFF circuits, which was given the moniker FortunaSPAD, was fabricated in the GF 55 nm BCD process with the aim of achieving multi-gigabit operation without the need of post-processing. The block diagram is shown in \figurename~\ref{fig:fortunablock}. FortunaSPAD contains two independent sub-arrays of QRFFs that can be operated simultaneously, along with readout and control circuitry. The chip micrograph and system testing infrastructure is illustrated by \figurename~\ref{fig:chiparchandtest}. A requirement of the system was to service two different interfaces, which is the reasoning for the two separate arrays. 

The first sub-array, denoted as $\mathrm{A1}$, contains 70x32 QRFFs, which are individually read-out through an output multiplexer. Furthermore, in this sub array each individual pixel is combined with a multiplexer controlled by $\mathrm{COUNT}$ (\figurename\ref{fig:pix}), which can bypass the TFF/DFF circuit, allowing for monitoring of the count rate. This enables a comparison between expected results, based on the model, and measurements, along with a more quantitative method for which to decide the illumination intensity.

The second sub-array, $\mathrm{A2}$, contains a more complex readout scheme. An on-chip digital PLL is used to operate a serializer block which serializes 70 SPADs onto a single readout channel. In order to ensure that data transmitting from the chip to the FPGA is valid, the FortunaSPAD contains a control flag that, when enabled, outputs a known pattern to the FPGA. The FPGA is then able to tune the IO delays of each channel appropriately until the known pattern is received.

The two sub-arrays are read out to two separate FPGAs for firmware simplicity, although there is nothing precluding the system from using a single FPGA. A motherboard containing all the required voltage generation and illumination control for the ASIC is designed so that the entire QRNG can be operated using a USB interface. An optical tube houses the LED and a diffuser in order to provide a uniform illumination across the array while also shielding external light. The LED wavelength is 470 nm, which was chosen based on measurements of the photon detection probability (PDP), described in the following section. The FortunaSPAD die area is 1.72 x 2.1 mm with horizontal and vertical pixel pitches of 24 $\mu$m and 35 $\mu$m, respectively. 

\section{Measurement Results}
\subsection{SPAD performance characterization}
\subsubsection{Specifications} 
\begin{figure}[!ht]
    \centering
    \includegraphics[]{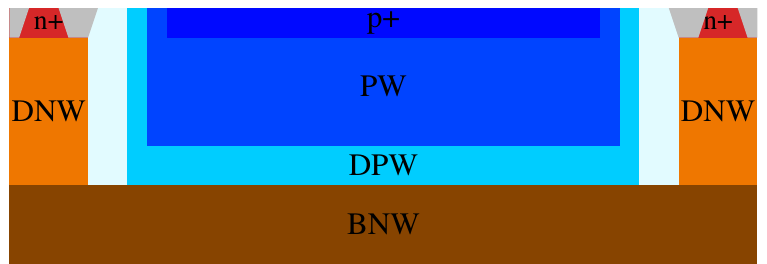}
    \caption{Cross-section of 55 nm BCD SPAD used in the FortunaSPAD. The junction is formed by the DPW/BNW interface.}
    \label{fig:pin3cross}
\end{figure}

The design of the SPAD is similar to that published in \cite{GramugliaKesharzianKizilkan_Eng55nm_JSTQE2022} with the cross-section shown in \figurename~\ref{fig:pin3cross}. The junction is buried deep inside the silicon using a deep p-well, buried n-well (DPW/BNW) implants. The advantage of using a deep junction is that they typically have lower afterpulsing, compared to shallower junctions, as traps from the silicon oxide interface have a greater distance to diffuse in order to enter the multiplication region and cause a spurious avalanche. Furthermore, the PDP is enhanced, enabling a larger spectrum from which to choose the illumination wavelength. The SPAD active radius is 4.4 $\mu$m, a virtual guard ring spanning 1 $\mu$m on each side and a total radius of 6.5 $\mu$m. 
\subsubsection{Afterpulsing} 
\begin{figure}[!t]
    \centering
    \includegraphics[]{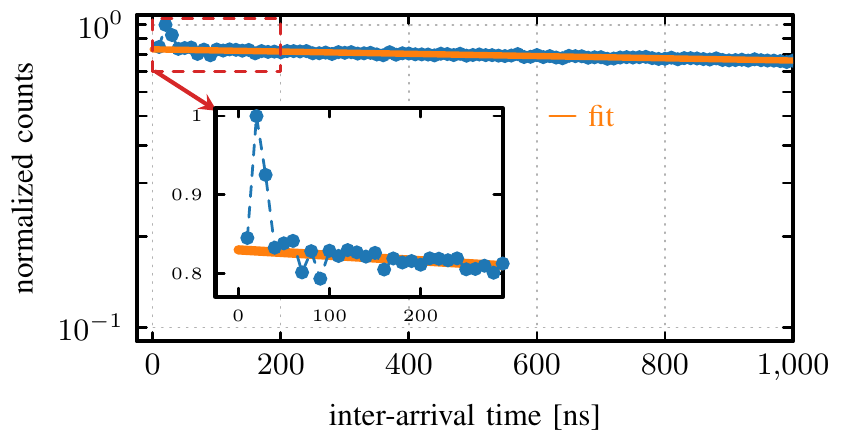}
    \caption{Afterpulsing measurement performed at room temperature using the inter-arrival histogramming method.}
    \label{fig:ppixelaft}
\end{figure}

As discussed, perhaps the most critical parameter of the SPAD is afterpulsing, as it induces correlated noise into the random bit generation circuitry. Using the inter-arrival time histogramming technique we estimate the afterpulsing by connecting the test pixel output to a fast 40 GS/s oscilloscope (Teledyne LeCroy WaveMaster 813 Zi-B) with an active probe and bin width of 10 ns. The pixel dead time was tuned to $\approx$ 8 ns, in order to attain accurate measurements for high-count rate applications. A low-level of light was added to the measurement, in order to attain a count rate $\approx$ 1 kcps. The results of the experiment are shown in \figurename~\ref{fig:ppixelaft}. The extracted afterpulsing is $\approx$ 0.005 \%. From the histogram, it can be seen that the lifetime of traps decays completely after approximately 100 ns. Both the lifetime and afterpulsing percentage are excellent results for a silicon SPAD in a deep sub-micron process. 
\subsubsection{PDP} 
\begin{figure}[!ht]
    \centering
    \includegraphics[]{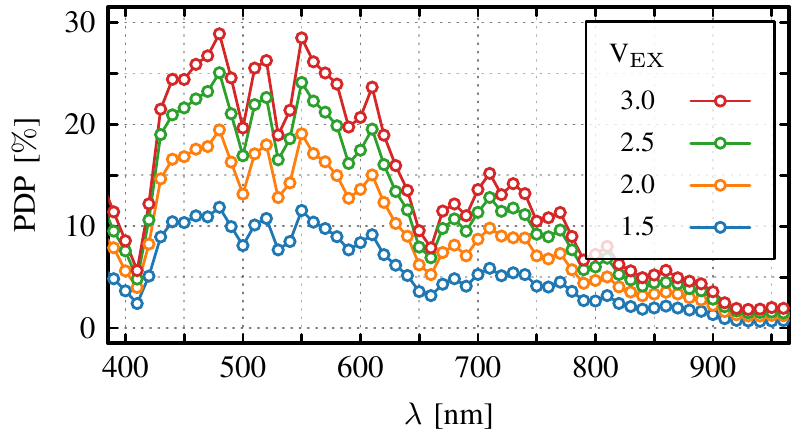}
    \caption{PDP measured using integrated PQAR circuit at room temperature across excess bias.}
    \label{fig:pin3PDP}
\end{figure}
The same test pixel was used for measurement of the PDP, with results shown in \figurename~\ref{fig:pin3PDP}. The data was taken using the continuous light method at 10 nm intervals up to 3 volts excess bias ($\mathrm{V_{EX}}$) using a setup that has been detailed in \cite{Francesco12ps:article_typical}. Due to the process, which was not optimized for image sensing, a clear standing wave pattern is seen across the spectrum. An LED (Cree C503B-BAN-CZ0A0452) in the blue spectrum ($\lambda= 470$ nm) is selected for the QRNG in order to avoid the efficiency troughs caused by this standing wave pattern, while maintaining a high relative detection efficiency to avoid using higher LED current. 

\subsubsection{DCR} 
\begin{figure}[!ht]
    \centering
    \includegraphics[]{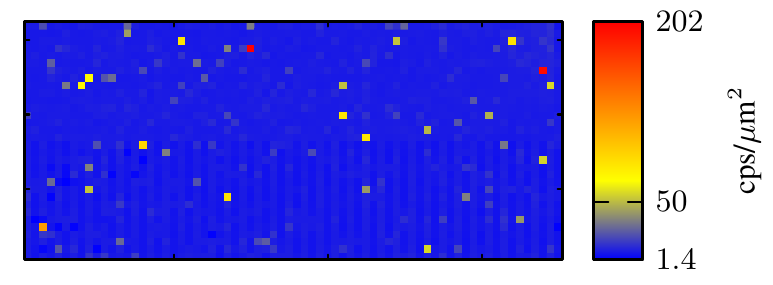}
    \caption{Normalized DCR of each pixel in $\mathrm{A1}$ array measured at room temperature.}
    \label{fig:DCRnorm}
\end{figure}

The dark count rate was measured across all pixels in $\mathrm{A1}$, by bypassing the random flip-flop circuitry. The results are shown in \figurename~\ref{fig:DCRnorm}. The DCR across all pixels remains relatively low with 95 \% of pixels remaining $<$10 cps$/\mu$m$^2$ with only three `hot' pixels that are $>$100 cps$/\mu$m$^2$. Therefore, all QRFF's in $\mathrm{A1}$ should be operable in the desired entropy bounds if circuit and illumination parameters are chosen carefully.
\subsubsection{Counting}
\begin{figure}[!ht]
    \centering
    \includegraphics[]{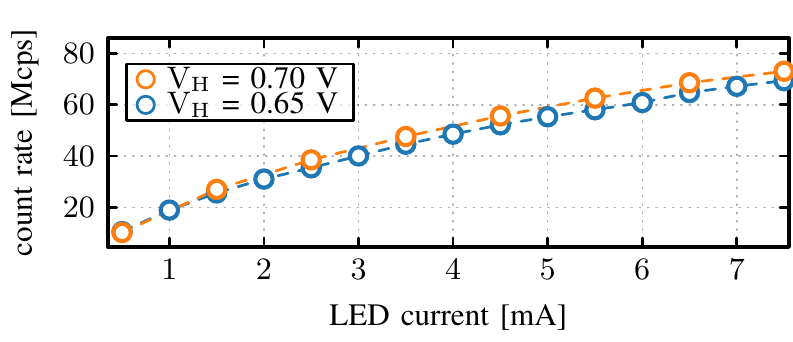}
    \caption{Count sweep of single test pixel with swept LED current measured at room temperature.}
    \label{fig:iledcountsweep}
\end{figure} 
As an initial validation of the model and to observe the performance capabilities of a single QRFF, the count rate is measured across swept led current with the results shown in \figurename~\ref{fig:iledcountsweep}. The measurements are taken with two different control voltages for the hold time with $\mathrm{V_H}=0.65$ V and $\mathrm{V_H}=0.70$ V resulting in dead times of $\approx$ 10 ns and $\approx$ 8 ns, respectively. Increasing $\mathrm{V_H}$ i.e. decreasing the dead-time past 0.70 V causes the pulse-width to shrink to a level where the count rate is not consistently measurable. Nevertheless, the results show counting performance that increases almost linearly with led current with perhaps some pile-up observed for $\mathrm{I_{LED}} > 2.0$ mA at $\mathrm{V_H}=0.65$ V. The LED itself can also be a source of non-linearity.      
\begin{figure*}[!ht]
\centering
\subfloat[1,2, and 3-bit lag correlation coefficients with swept sampling frequency, $f_{\mathrm{BG}}$, and $\mathrm{I_{LED}}=5$ mA. ]{
\includegraphics[]{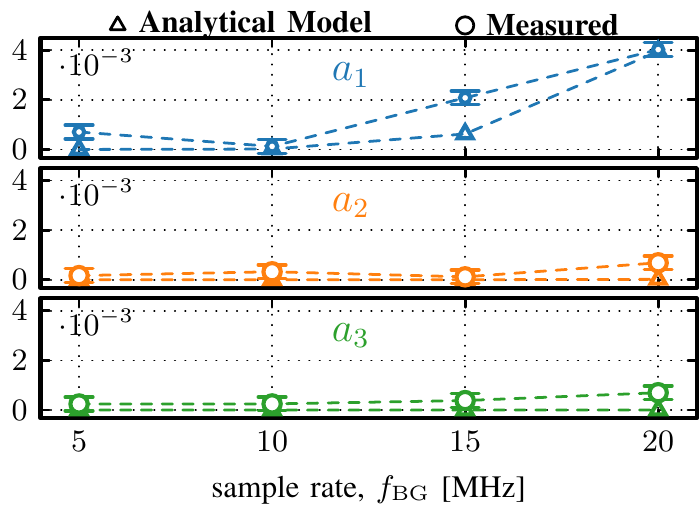}
\label{fig:pix902a1-3}}\hfil
\subfloat[$P=1$ bias at $f_{\mathrm{BG}}=5$ MHz as a function of normalized sampling threshold and LED current.]{
\includegraphics[]{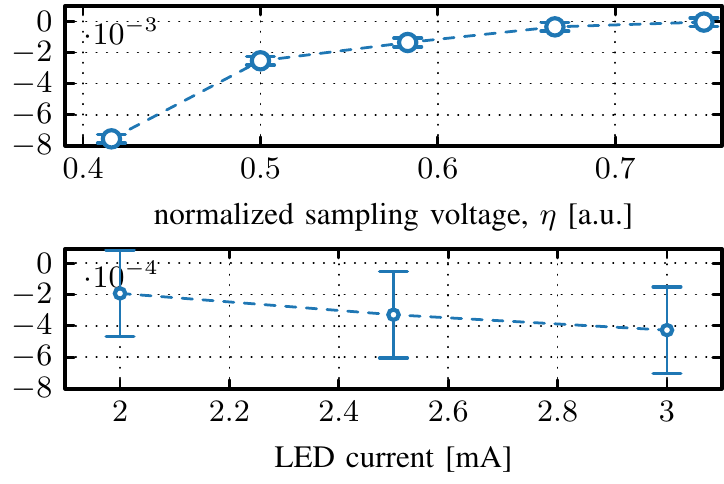}
\label{fig:pix902iledbias}}
\caption{Measured bias and autocorrelation results of a test pixel for comparison to expected results based on derived analytical model. Measurements performed with $\mathrm{V_{H}}=0.7$ V.}
\label{fig:pix902analvsmeas}
\end{figure*}
\subsection{Comparison between analytical and measurement of a single QRFF}
The performance of a single QRFF was evaluated for comparison with the expected results. Results for correlation and bias can be seen in \figurename~\ref{fig:pix902analvsmeas}. Correlation coefficients are compared to the analytical values, since they are a function of measurable qualities ($\lambda_\mathrm{D}$, $f_\mathrm{BG}$). They remain at acceptable values up until $f_\mathrm{BG}=15$ MHz under these operating conditions and, as expected, higher-order coefficients become non-negligible only at high sampling rates. The bias can be seen to scale linearly with increased illumination (count rate). This is observed at lower illumination values in order to avoid pile-up. Moreover, we can see that the critical hypothesis regarding sampling threshold is confirmed. By adjusting the sampling threshold of the QRFF we are able to essentially eliminate bias by balancing the mismatch in the TFF output waveform. 

In order to test the limits on performance of the QRFF, the dead-time is reduced to a minimum by adjusting $\mathrm{V_H}$ to $\mathrm{VDD}$ and keeping the value of $\mathrm{V_R}$ to a low value of 0.1 V in order to avoid any effects from pile up. A summary of results for correlation and bias are shown in Table \ref{tab:qrffpix902a1bias}. The results for both bias and correlation remain above the acceptable entropy bound even until 25 MHz. These results were calculated by generating 327 Mb of data for each sample rate, therefore the calculated $\sigma$ for bias and correlation are 2.76E$^{-5}$, and 1.3E$^{-5}$, respectively.
\begin{table}[!ht]\renewcommand{\arraystretch}{1.2}
\caption{Single-QRFF Minimum Dead-Time Results}
\centering
\begin{tabular}{|c|c||c|c|c|}
\hline
\textbf{$\mathrm{f_{BG}}$} &$\mathrm{I_{LED}}$ [mA]& $a_1$&$b$\\ \hline\hline
5 &5 & 6.58E$^{-5}$ & 1.76$^{-5}$ \\ \hline 
10 &5 & -4.97$^{-5}$& 2.34$^{-4}$ \\ \hline 
15 &5 & 1.51E$^{-4}$ & 3.56E$^{-4}$ \\ \hline 
20 &5 & 4.66E$^{-4}$ & 3.23E$^{-4}$ \\ \hline 
25 &5 & 8.32E$^{-4}$ & 2.39E$^{-4}$ \\ \hline 
30 &6.5 & 1.45E$^{-3}$ & 4.21E$^{-4}$ \\ \hline 
\end{tabular}\label{tab:qrffpix902a1bias}
\end{table}\renewcommand{\arraystretch}{1.0}

\subsection{Array performance characterization}
\subsubsection{SPAD operating voltage}
\begin{figure*}[!ht]
\centering
\subfloat[$\mathrm{V_{OP}=32.8}$ V.]{
\includegraphics[]{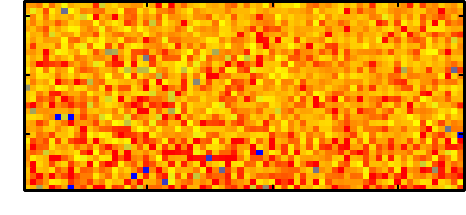}
\label{fig:vopbias32p8}}
\subfloat[$\mathrm{V_{OP}=33.1}$ V.]{
\includegraphics[]{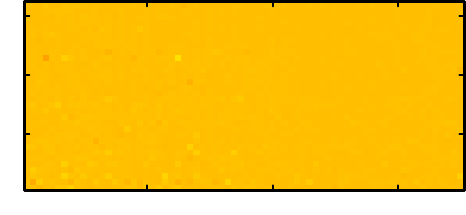}
\label{fig:vopbias33p1}}\hspace{-8mm}
\subfloat[$\mathrm{V_{OP}=33.3}$ V.]{
\includegraphics[]{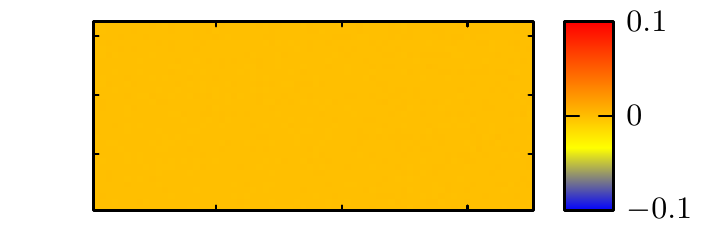}
\label{fig:vopbias33p3}}
\caption{Spatial bias map across excess bias with constant illumination, $\mathrm{I_{LED}}=2$ mA and a sampling rate $f_{\mathrm{BG}}=5$ MHz.}
\label{fig:vopbiasspatial}
\end{figure*}
In order to determine proper operation of the chip, the non-uniformity of breakdown voltages across the array must be understood. The $\mathrm{V_{OP}}$ should then be set to the minimum value of excess bias where all QRFFs are operating correctly, in order to reduce effects of afterpulsing. A method that can be used to determine this voltage is to observe the per QRFF bit bias at a constant illumination while increasing excess voltages. A visualization of the results from this test is shown in \figurename~\ref{fig:vopbiasspatial}, where a spatial heat map of the per QRFF bias is shown. It can be observed that as the excess voltage is increased, the bit bias reaches a uniform (low) value, at a $\mathrm{V_{OP}}=33.3$ V, which is the operating value used for all subsequent measurements.   

\subsubsection{Bias and correlation analysis as function of model parameters}
\begin{figure*}[!ht]
\centering
\subfloat[RMS bias.]{
\includegraphics[]{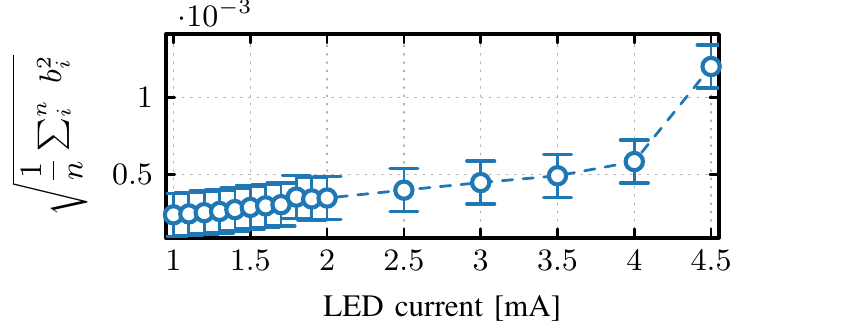}
\label{fig:bias_rms_iledsweep}}
\subfloat[Mean bias.]{
\includegraphics[]{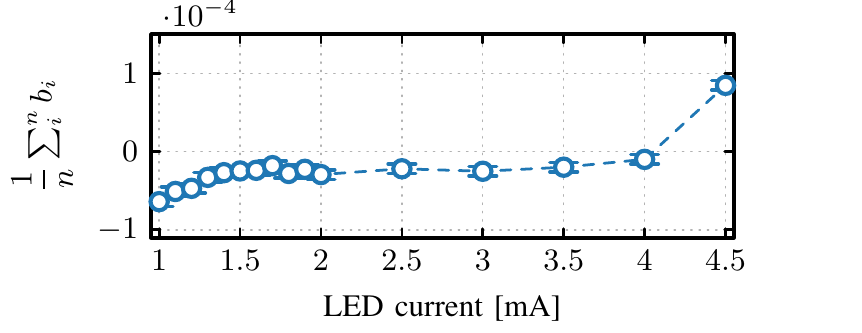}
\label{fig:bias_mean_iledsweep}}
\\
\subfloat[RMS 1-bit lag autocorrelation coefficient.]{
\includegraphics[]{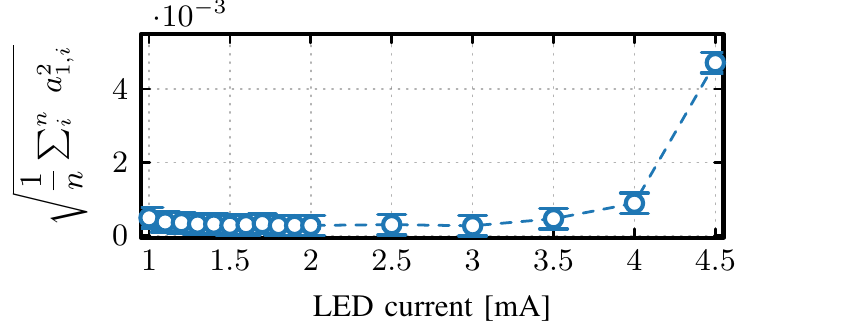}
\label{fig:a1corrrmsiledsweep}}
\subfloat[Mean 1-bit lag autocorrelation coefficient.]{
\includegraphics[]{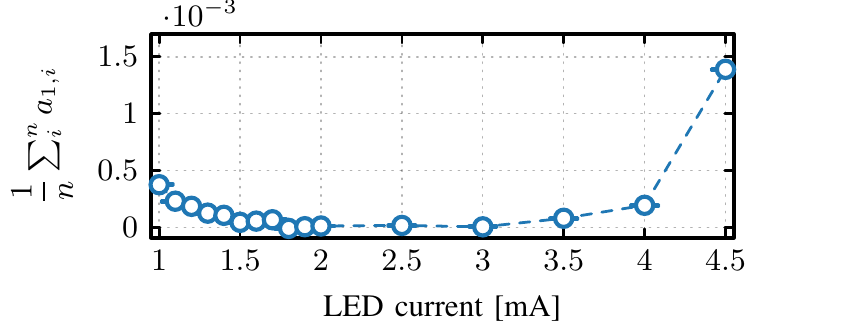}
\label{fig:a1corrmeaniledsweep}}
\caption{Bias and correlation analysis across all QRFFs in A1 ($n=2240$) as a function of LED current at a $f_{\mathrm{BG}}=5$ MHz.}
\label{fig:iledsweep}
\end{figure*}
The root mean square (RMS) and mean values of per QRFF bias, $b$, and serial correlation, $a_1$, are shown as a function of illumination intensity in \figurename~\ref{fig:iledsweep} with $f_{\mathrm{BG}}= 5$ MHz. From the perspective of bit bias, the RMS value across the array increases with an increase in LED current, as expected, since the higher count rates scale bias proportionally. Meanwhile, it is observed that the mean bias from 2-3.5 mA remains constant, as a constant sampling threshold ($\eta$) is maintained for all tests. A deviation from this constant magnitude of the mean bias between 1-1.5 mA is observed. This is caused by a low number of pixels, which remain at lower count rates, therefore shifting the mean of the bias slightly. 

\begin{figure}[!ht]
\centering
\subfloat[Mean bias.]{
\includegraphics[]{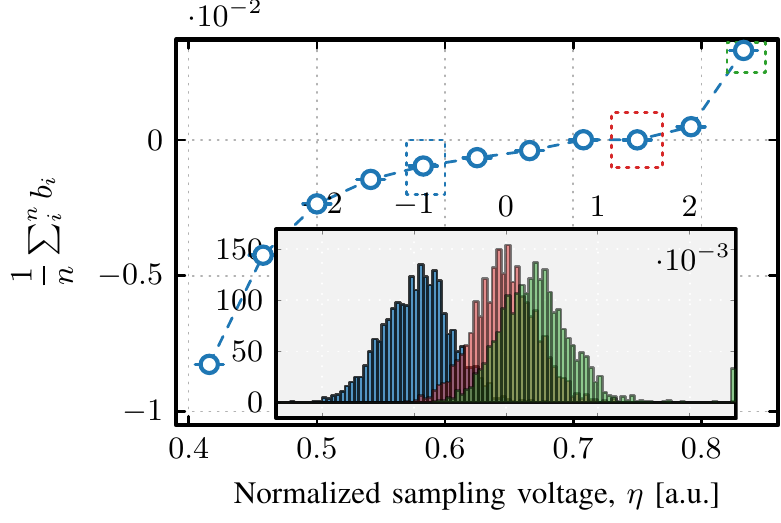}
\label{fig:bias_mean_vtsweep}}\\
\subfloat[Mean 1-bit lag autocorrelation coefficient.]{
\includegraphics[]{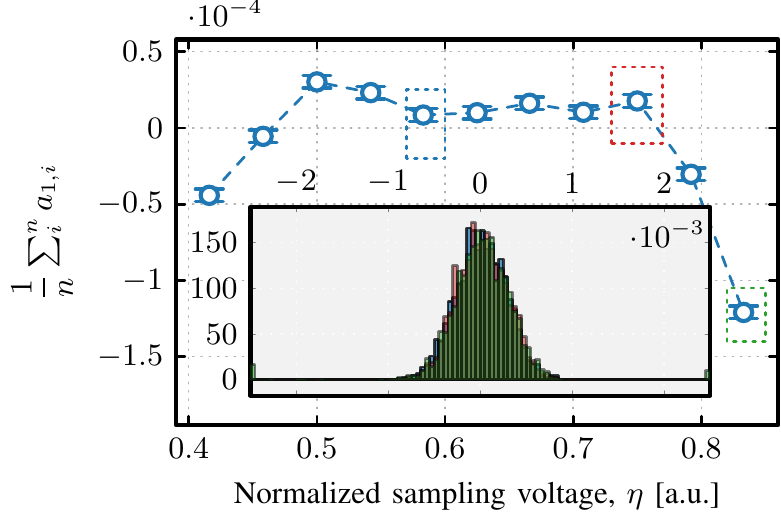}
\label{fig:a1_mean_vtsweep}}
\caption{Per QRFF analysis across A1 array at a 5 MHz bit generation rate with swept sampling threshold voltage.}
\label{fig:vtsweep}
\end{figure}

The sampling threshold is also swept and in doing so, the mean bias of the entire array is very close to 0. The results are shown in \figurename\ref{fig:vtsweep}. Three points along the curve are also placed in a histogram to visualize the shifting of the entire array in bias, while remaining unchanged for autocorrection. At higher values for the sampling threshold, a small amount of pixels becomes stuck, as their inherent comparator offset prevents the toggling of the output.

\subsection{A2 Performance}
\begin{figure}[!ht]
\centering
\subfloat[$b$.]{
\includegraphics[]{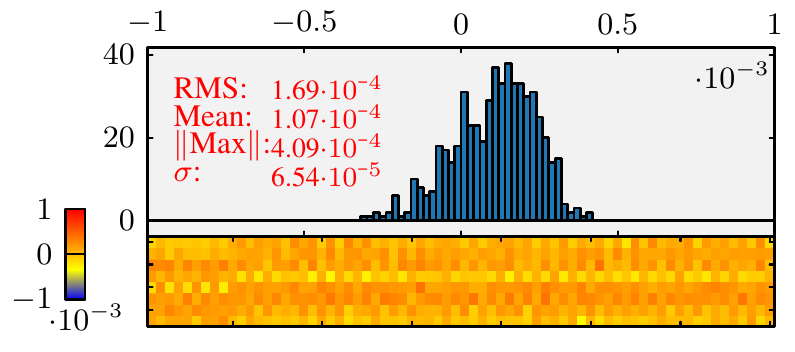}
\label{fig:hs_2mA_5MHz_b}}\\
\subfloat[$a_1$.]{
\includegraphics[]{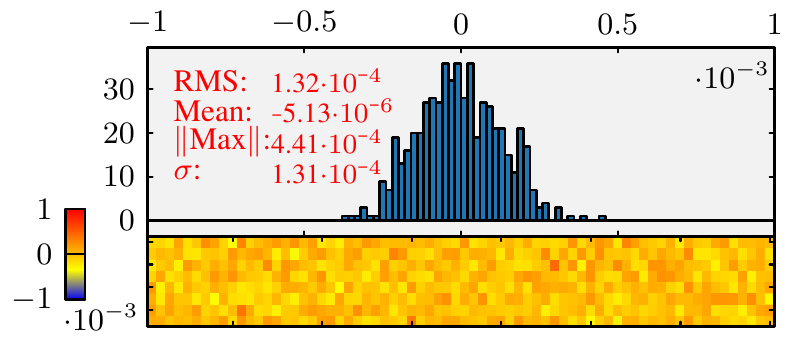}
\label{fig:hs_2mA_5MHz_a1}}
\caption{A2 spatial maps at $f_\mathrm{BG}= 15$ MHz and $\mathrm{I_{LED}}=2$ mA}
\label{fig:a2spatialmaps}
\end{figure}

In order to characterize the serialized array, a strobe signal is also implemented inside the FortunaSPAD, which is synchronized to the first QRFF output in the array. This allows for a per pixel spatial analysis in order to make sure there are no malfunctioning circuits/detectors and no particular `hot' spots in the array. The calculated bias and correlation coefficient of all QRFFs in A2 are shown in \figurename\ref{fig:a2spatialmaps}. All QRFF's in the serialized array achieve a bias and serial correlation coefficient within the required entropy bounds. The max calculated bias and correlation are, 4.09x$10^{-4}$ and 4.41x$10^{-4}$, respectively, with RMS values across the array of 1.69x$10^{-4}$ and 1.32x$10^{-4}$, respectively. 

Overall for both arrays, under the same operating conditions, only 4 pixels fall slightly outside these benchmarks. Although the 2796 QRFFs within the entropy bounds are capable of generating 14 Gbps of data, the limitations of the readout circuitry and IOs results in a combined data rate of 3.3 Gbps. In order to ensure that no spatial cross-correlations affect the results of the generated bit strings, a full frame of data is readout in a single $\mathrm{CLK_{BG}}$ cycle for statistical testing.   
\subsection{NIST-STS}
The ability to achieve erroneous results from the NIST Statistical Test Suite when incorrect parameters are chosen is well documented \cite{Kinga_NISTinterp_SciAndTech2015,HurleySmith_QLeapCreash_ACM2020}. Therefore, we choose strict parameters for NIST testing with 1 Gb of data split into 1000 bit strings using a significance level ($\alpha$) of 0.001. The results for the NIST test are outlined in Table~\ref{tab:NISTresults} with all tests passing. 
\begin{table}[!ht]
\begin{center}
\scalebox{1}{\begin{threeparttable}
\renewcommand{\arraystretch}{1.1}
\caption{Sample Summary of NIST Results}
\label{tab:NISTresults}

\begin{tabular}{c|c|c|c|c|c|c|c}
\hline
\textbf{Test}
& \textbf{\begin{tabular}[c]{@{}c@{}} Min. pass rate \end{tabular}} 
& \textbf{\begin{tabular}[c]{@{}c@{}} $p$-value \end{tabular}}
& \textbf{\begin{tabular}[c]{@{}c@{}} Pass rate \end{tabular}} 
\\ \hline\hline

\multicolumn{1}{l|}{Frequency} 
& \begin{tabular}[c]{@{}c@{}} 996\end{tabular}       
&\begin{tabular}[c]{@{}c@{}}  0.8831\end{tabular}     
& \multicolumn{1}{c}{998/1000}
\\ \hline
\multicolumn{1}{l|}{Block frequency} 
& \begin{tabular}[c]{@{}c@{}} 996\end{tabular}       
&\begin{tabular}[c]{@{}c@{}}  0.0278\end{tabular}     
& \multicolumn{1}{c}{1000/1000}
\\ \hline

\multicolumn{1}{l|}{Cumulative sums} 
& \begin{tabular}[c]{@{}c@{}} 996\end{tabular}       
&\begin{tabular}[c]{@{}c@{}}  0.1855\end{tabular}     
& \multicolumn{1}{c}{997/1000}
\\ \hline
\multicolumn{1}{l|}{Runs} 
& \begin{tabular}[c]{@{}c@{}} 996\end{tabular}       
&\begin{tabular}[c]{@{}c@{}}  0.4521\end{tabular}     
& \multicolumn{1}{c}{999/1000}
\\ \hline
\multicolumn{1}{l|}{Longest run} 
& \begin{tabular}[c]{@{}c@{}} 996\end{tabular}       
&\begin{tabular}[c]{@{}c@{}}  0.4885\end{tabular}     
& \multicolumn{1}{c}{998/1000}
\\ \hline
\multicolumn{1}{l|}{Rank} 
& \begin{tabular}[c]{@{}c@{}} 996\end{tabular}       
&\begin{tabular}[c]{@{}c@{}}  0.9723\end{tabular}     
& \multicolumn{1}{c}{998/1000}
\\ \hline
\multicolumn{1}{l|}{FFT} 
& \begin{tabular}[c]{@{}c@{}} 996\end{tabular}       
&\begin{tabular}[c]{@{}c@{}}  0.1364\end{tabular}     
& \multicolumn{1}{c}{999/1000}
\\ \hline
\multicolumn{1}{l|}{Non overlapping template} 
& \begin{tabular}[c]{@{}c@{}} 996\end{tabular}       
&\begin{tabular}[c]{@{}c@{}}  0.8429\end{tabular}     
& \multicolumn{1}{c}{997/1000}
\\ \hline
\multicolumn{1}{l|}{Overlapping template} 
& \begin{tabular}[c]{@{}c@{}} 996\end{tabular}       
&\begin{tabular}[c]{@{}c@{}}  0.6454\end{tabular}     
& \multicolumn{1}{c}{998/1000}
\\ \hline
\multicolumn{1}{l|}{Universal} 
& \begin{tabular}[c]{@{}c@{}} 996\end{tabular}       
&\begin{tabular}[c]{@{}c@{}}  0.7830\end{tabular}     
& \multicolumn{1}{c}{1000/1000}
\\ \hline
\multicolumn{1}{l|}{Approximate entropy} 
& \begin{tabular}[c]{@{}c@{}} 996\end{tabular}       
&\begin{tabular}[c]{@{}c@{}}  0.5769\end{tabular}     
& \multicolumn{1}{c}{1000/1000}
\\ \hline
\multicolumn{1}{l|}{Random excursions} 
& \begin{tabular}[c]{@{}c@{}} 616\end{tabular}       
&\begin{tabular}[c]{@{}c@{}}  0.3258\end{tabular}     
& \multicolumn{1}{c}{618/619}
\\ \hline
\multicolumn{1}{l|}{Random excursions variant} 
& \begin{tabular}[c]{@{}c@{}} 616\end{tabular}       
&\begin{tabular}[c]{@{}c@{}}  0.5457\end{tabular}     
& \multicolumn{1}{c}{616/619}
\\ \hline
\multicolumn{1}{l|}{Serial} 
& \begin{tabular}[c]{@{}c@{}} 996\end{tabular}       
&\begin{tabular}[c]{@{}c@{}}  0.9737\end{tabular}     
& \multicolumn{1}{c}{1000/1000}
\\ \hline
\multicolumn{1}{l|}{Linear complexity} 
& \begin{tabular}[c]{@{}c@{}} 996\end{tabular}       
&\begin{tabular}[c]{@{}c@{}}  0.5523\end{tabular}     
& \multicolumn{1}{c}{999/1000}
\\ \hline

\end{tabular}
\end{threeparttable}}

\end{center}
\end{table}\renewcommand{\arraystretch}{1}

\section{Discussion and Comparison}
A summary of relevant integrated SPAD-based QRNGs, which include the bit generation/extraction method on-chip is shown in Table \ref{tab:SPAD_QRNGs}. It can be seen that for an SPAD array-based solution with bit generation on chip, we demonstrate the highest per-pixel generation rate reported. Furthermore, the ability of a single pixel to generate 25 Mbps is the highest reported for an integrated solution. Most prior art either rely on the quantum nature of the entropy source or an arbitrarily chosen post-processing method for justification of the bit generation quality. However, in our work, we systematically model the degradation of entropy and validate it through simulation. As a result, we were able to propose a circuit innovation which was capable of overcoming this, without the expense of a reduced generator speed, an outcome that would inevitably be the case if post-processing was employed.  

\begin{table*}[!t]
\begin{center}
\scalebox{0.9}{\begin{threeparttable}
\renewcommand{\arraystretch}{1.3}
\caption{Published Integrated SPAD-Based QRNGs}
\label{tab:SPAD_QRNGs}

\begin{tabular}{l|c|c|c|c|c|c}
\hline\hline
\textbf{Ref. \& Year}
& \textbf{\begin{tabular}[c]{@{}c@{}} Physical\\Principle \end{tabular}} 
& \textbf{\begin{tabular}[c]{@{}c@{}} Circuit/Extraction\\Implementation \end{tabular}}
& \textbf{\begin{tabular}[c]{@{}c@{}} Array\\Size \end{tabular}} 
& \textbf{\begin{tabular}[c]{@{}c@{}} Bitrate\\(per pixel) \end{tabular}}
& \textbf{\begin{tabular}[c]{@{}c@{}} Further Post\\Processing \end{tabular}}  
& \textbf{\begin{tabular}[c]{@{}c@{}} Evaluation\\Method \end{tabular}} 
\\ \hline\hline

\multicolumn{1}{l|}{\cite{regazzoni_qrng_2021} Regazzoni, 2021} 
& \begin{tabular}[c]{@{}c@{}}Photon detection \end{tabular}
& \begin{tabular}[c]{@{}c@{}} Vector matrix\\ multiplication\tnote{$\dagger$} \end{tabular}
& \begin{tabular}[c]{@{}c@{}}128x128$\dagger$\end{tabular} 
& \begin{tabular}[c]{@{}c@{}} 400 Mbps\\ ($\simeq$~0.024 Mbps)\end{tabular}
& \begin{tabular}[c]{@{}c@{}}none\end{tabular}
& \begin{tabular}[c]{@{}c@{}}NIST STS\\ Diehard\end{tabular}
\\ \hline


\multicolumn{1}{l|}{\cite{Acerbi_emitterQRNG_JSTQE2018} Acerbi, 2018.} 
& \begin{tabular}[c]{@{}c@{}}SPAD triggering\\probability \end{tabular}
& Frame readout
& 1 
& $\simeq$~0.1 Mbps
& \begin{tabular}[c]{@{}c@{}}none\end{tabular}
& \begin{tabular}[c]{@{}c@{}}NIST STS\end{tabular}
\\ \hline

\multicolumn{1}{l|}{\cite{Xu_aribterQRNG_TCASII2018} Xu, 2018.} 
& \begin{tabular}[c]{@{}c@{}}First detected\\photon \end{tabular}
& \begin{tabular}[c]{@{}c@{}} Inter-arrival\\ arbiter \end{tabular}
&  16x16 
& \begin{tabular}[c]{@{}c@{}} 18 Mbps\\ ($\simeq$~0.07 Mbps) \end{tabular}
& none
& NIST STS
\\ \hline

\multicolumn{1}{l|}{\cite{Massari_arbiterQRNG_ISSCC2016} Massari, 2016.} 
& \begin{tabular}[c]{@{}c@{}}First detected\\photon \end{tabular}
& \begin{tabular}[c]{@{}c@{}} Inter-arrival\\ arbiter \end{tabular}
&  16x16 
& \begin{tabular}[c]{@{}c@{}} 128 Mbps\\ (0.5 Mbps) \end{tabular}
& none
& NIST STS 
\\ \hline


\multicolumn{1}{l|}{\cite{Acerbi_emitterQRNG_JSTQE2018} Tisa, 2015.} 
& \begin{tabular}[c]{@{}c@{}}SPAD triggering\\probability \end{tabular}
& LFSR counter
& 32x32 
& \begin{tabular}[c]{@{}c@{}} 200 Mbps\\ ($\simeq$~0.2 Mbps)\end{tabular}
& \begin{tabular}[c]{@{}c@{}}Whitening\\algorithm\end{tabular}
& \begin{tabular}[c]{@{}c@{}}DIEHARDER\\TestU01\end{tabular}
\\ \hline

\multicolumn{1}{l|}{\cite{Burri_Jailbreak_ISSW2013} Burri, 2013.} 
& \begin{tabular}[c]{@{}c@{}}SPAD triggering\\probability \end{tabular}
& Frame readout
& \begin{tabular}[c]{@{}c@{}}512x128 \\ x2\tnote{$\alpha$}\end{tabular} 
& \begin{tabular}[c]{@{}c@{}} 5 Gbps\\ ($\simeq$~0.04 Mbps)\end{tabular}
& \begin{tabular}[c]{@{}c@{}}Von\\ Neumann \\ filter\end{tabular}
& \begin{tabular}[c]{@{}c@{}}NIST STS \\ DIEHARD\end{tabular}
\\ \hline

\multicolumn{1}{l|}{\textbf{This work}} 
& \begin{tabular}[c]{@{}c@{}}Photon timing\\ statistics \end{tabular}
& QRFF
& \begin{tabular}[c]{@{}c@{}}40 x 70$\ddagger$\end{tabular} 
& \begin{tabular}[c]{@{}c@{}} 3.3 Gbps\\ ($\simeq$~1.2 Mbps)\tnote{$\bowtie$}\\(25 Mbps)\tnote{*}\end{tabular}
& \begin{tabular}[c]{@{}c@{}}none\end{tabular}
& \begin{tabular}[c]{@{}c@{}}NIST STS\end{tabular}
\\ \hline

\end{tabular}
\begin{tablenotes}
\item[$\dagger$] a fixed matrix and reconfigurable matrix are included on chip
\item[$\ddagger$] two independent arrays (70 x 32 \& 70 x 8) with different readout architectures
\item[$\bowtie$] calculated per pixel throughput with both arrays readout, limited by readout and IO speeds
\item[*] single pixel capable of Shannon entropy > 0.997
\item[$\alpha$] Two of the same die were used in parallel.
\end{tablenotes}
\end{threeparttable}}

\end{center}
\end{table*}\renewcommand{\arraystretch}{1}
\section{Conclusion}
We have demonstrated a full multi-Gbps integrated SPAD-based QRNG system when using external illumination based on the QRFF method. The QRFF is an architecturally simple but feature-rich, scalable, model-testable bit generation method. By analyzing the degradation of entropy caused by circuit limitations, we were able to propose and validate a simple circuit innovation, namely the addition of a tunable sampling threshold, in order to essentially eliminate bias from a single QRFF. This opens the door for more complex QRNG systems based on our circuit technique, that can continually monitor and correct for changes in operation caused by, for example, changes in environmental settings. Furthermore, the ability to precisely control the generator bias and correlation is interesting for certain applications, such as stochastic computing \cite{alaghi_stochasCompProm_TCAD2017}.

To the authors' knowledge, the total throughput of 3.3 Gbps is the highest reported for a single-die SPAD-based system that also integrates its bit generation circuitry. Moreover, the generation capability of a single QRFF of 25 Mbps while maintaining a Shannon entropy > 0.997 is the highest single pixel throughput reported. We have taped-out an improved version of the FortunaSPAD, which contains an improved readout method for higher throughput along with integrated illumination. 

\bstctlcite{IEEEexample:BSTcontrol}
\IEEEpeerreviewmaketitle

\section*{Acknowledgment}
The authors of this paper would like to thank all of the members of the Beryllium project team who have contributed valuable input in discussions during development. 
\ifCLASSOPTIONcaptionsoff
  \newpage
\fi



\bibliographystyle{IEEEtran}
\bibliography{IEEEabrv,jsscfortuna_bib}

\end{document}